\documentclass{article}

\title{Eigenvalues of the static, spherically-symmetric Einstein-Proca equations}

\author{Chris Vuille \\ Department of Physical Sciences \\ Embry-Riddle Aeronautical University \\ Daytona Beach, FL 32114}

\begin{document}

\maketitle
\begin{abstract} The Proca potential has been used in many contexts in flat spacetime, but not so often in curved spacetime. Here, the static Proca potentials are derived on a Schwarzschild background, and are found to have very different forms from those in flat space. In addition, the coupling of the two fields leads to a quantum condition on the particle mass, which is found to be quantized and inversely proportional to the range parameter, $\mu$. This suggests that gravity may induce quantization in other physical fields. 
\end{abstract}

\pagebreak

\section{Introduction}
	The Proca equation was originally investigated by Proca \cite{Proca1} \cite{Proca2}.  It has been successfully used in field theories for some decades \cite{Poenaru1}-\cite{Poenaru}, originating as a generalization of Maxwell's equations to a massive spin-1 field.  In nuclear theory, it has been used in connection with the strong force \cite{Pauli}, and it has also been cited as a possible source for dark matter and a partial explanation for the galactic rotation curves, according to Tucker and Wang \cite{Tucker}. Some general development of the Einstein-Proca system was undertaken in Dereli et al. \cite{Dereli}. Vuille et al., \cite{Vuille} developed perturbative calculations  and made a case for the absence of event horizons for micro black holes, a possible counterexample to weak cosmic censorship.  Obukov and Vlachynsky \cite{Obukov}and Toussaint \cite{Toussaint} independently found numerical solutions. These latter two papers, confirmed by the work in \cite{Vuille}, demonstrated the existence of naked singularities in this system. In the context of an approximate solution for a modified field theory using ideas of Weyl and spacetimes with torsion, this problem was studied by Gottlieb et al. \cite{Gottlieb}, but their method missed the most interesting aspect of the system: gravity can stimulate the quantization of the Proca potentials.
	
This paper is a continuation of \cite{Vuille}, extending the calculations developed there to obtain the forms of the Proca potential on a static, spherically-symmetric spacetime. Solving the zeroth order perturbation equations are equivalent to solving the Proca equations on a Schwarzschild background. In doing so, it's of great interest that there are an infinite number of quantized solutions, where the particle's mass is inversely proportional to the range parameter, $\mu$. In turn, $\mu$ is usually interpreted as proportional to the mass of the Proca exchange particle.

\section{The Einstein-Maxwell-Proca Equations}

     The equation for a particle exhibiting a spin-1 field with an exchange particle having mass is  Proca's equation \cite{Proca1}: 

\begin{equation}\label{a}
\partial_a F^{ab}+\mu^2 A^b= kj^a
\end{equation}
where the current $j^a$ will be taken to be zero in the case of vacuum, while 
\begin{equation}\label{b}
F_{ab}=\nabla_a A_b -\nabla_b A_a
\end{equation}
The quantity $\mu$ is  a constant, often interpreted as being proportional to the mass of the field quanta. It is associated with the rate of change of strength of the potential. The Lagrangian density for the Proca system in vacuum is \cite{Wald}:

\begin{equation}\label{h1}
\pounds=\sqrt{-g} \left(\alpha F_{ab}F^{ab}+ \beta A_a A^a\right)
\end{equation}
where  $g$ is the determinant of the metric and $\alpha$ and $\beta$ are constants.  Varying this
equation with respect to $A^c$ gives equation \ref{a}, provided that $\beta/2 \alpha=-\mu^2$ and the current is zero. 
For spherical symmetry, the metric has the form
\begin{equation}\label{l}
ds^2= e^\nu dt^2-e^{\lambda}dr^2-r^2 \left(d \theta^2+\sin^2 \theta d \phi^2 \right)
\end{equation}
The Proca stress-energy tensor can be obtained by varying the Lagrangian with respect to the metric:
\begin{equation}
T_{ab}=-\frac{\alpha_M}{8 \pi} \frac{1}{\sqrt{-g}} \frac{\delta \pounds}{\delta g^{ab}}
\end{equation}
The constant $\alpha_M$ is a parameter that must be adjusted to give the proper strength of the gravity field created by the stress-energy \cite{Wald}. It's considered very small, although that is a conjecture and may depend on scale. This constant and
the factor of $8 \pi$ will be combined with the constants $\alpha$ and $\beta$ for simplicity.  This calculation yields:
\begin{equation}\label{j}
T_{ab}=2 \alpha {F_a}^dF_{bd}+ \beta A_a A_b -\frac{1}{2}
g_{ab} \left( \alpha F_{cd}F^{cd}+ \beta A_c A^c \right)
\end{equation}
Einstein's equations are given by:
\begin{equation}\label{k}
R_{ab}=\kappa \left(T_{ab}-\frac{1}{2} Tg_{ab}\right)
\end{equation}

Following Carmeli \cite{Carmeli}, and Vuille et al. \cite{Vuille}, the full Einstein-Proca equations for static, spherical symmetry, with $f = e^\nu$, can be written as

\begin{equation} \label{18}
f''+\frac{2}{r} f' = -2 \kappa \alpha {A_0'}^2  +\kappa \beta
{A_0}^2 \left(2+\frac{r f'}{2f}\right) e^{\lambda}
\end{equation}

\begin{equation}
A_0'' + \frac{2}{r} A_0' =\frac{\beta}{2 \alpha} A_0 \left(-1+\frac{\alpha \kappa r A_0 A_0'}{f}\right) e^{\lambda} 
%
\end{equation}
where
\begin{equation} \label{20}
 e^{\lambda} = \left[ \frac{f+rf'-\kappa r^2 \left(\alpha {A_0'}^2\right)}{f +\frac{1}{2} \kappa \beta r^2 {A_0}^2}\right]
 \end{equation}

 \section{Proca Solution on a Schwarzschild Background}
 
There are no known solutions of the Einstein-Proca equations \cite{exact}. In Vuille et al.,  \cite{Vuille}, the zeroth order gravity field was taken to be flat space, resulting in a zeroth order  flat space Proca equation. The flat space Proca was then used to obtain a first-order solution for the gravity field.  The interest here, however, is discovering how the gravity field affects the solution for the Proca field. The base spacetime will therefore be taken to be the Schwarzschild spacetime. For details on developing the perturbation, the reader may consult Vuille et al. \cite{Vuille}. 

\subsection{First Solution}
In this section the Einstein-Proca equations will be solved exactly with the Schwarzschild solution as the background metric. This is essentially a zeroth order solution, with the zeroth order metric taken to be the Schwarzschild metric. The  equations are
\begin{equation}
f_0'' + \frac{2}{r} f_0 = 0
\end{equation}
and 
\begin{equation}
u_0'' + \frac{2}{r} u_0' = -p\mu^2 u_0 \left(1+\frac{rf_0'}{f_0}\right)
\end{equation}
where $p = \pm 1$, with $-1$ corresponding to the hyperbolic Proca and $+1$ the oscillatory Proca. Here, the focus will be on the hyperbolic Proca. it is evident that 
\begin{equation}
f_0 = 1 - \frac{A}{r}
\end{equation}
for some parameter $A$. Substituting the expression for $f_0$, then $u_0 = g/r$, and $ y = r-A$, obtain 
\begin{equation}
g'' - \mu^2g  - \mu^2 A \frac{g}{y} = 0
\end{equation}

This is related to Kummer's differential equation, the same equation as studied in \cite{Gottlieb} although they arrived at it as an approximation for a solution of a modified field theory. They rejected trial solutions involving a positive exponential as unphysical, being infinite at infinity, but it's important to realize that phenomenological quark potentials have that behavior. Further, they did not make the substitution $y = r-A$. The alternate analysis here yields new insight into the solution of this equation. Substituting  $g(z) = h_1(z) e^{z}$, where $z = \mu y$, yields

\begin{equation}
h_1'' + 2 h_1' - \frac{\mu A}{z} h_1 = 0
\end{equation}
where now the primes refer to derivatives with respect to $z$.
Following the technique in, for example, \cite{Schaums}, the point $z = 0$ is a singular point, and the characteristic polynomial has roots $\lambda_1 = 1$ and $\lambda_2 = 0$.
A solution can be found with the expansion

\[ h_1(z) = \sum_0^\infty a_n z^{n+1} \]
which corresponds to the root $\lambda_1 = 1$. It's straightforward to derive that for $n \ge 0$:

\begin{equation}
a_{n+1} = \left(\frac{\mu A - 2(n+1)}{(n+1)(n+2)}\right)a_n
\end{equation}
By the ratio test, this series will converge for all real numbers, as will the full solution, with the factor $e^z/z$, so there is a continuum solution for all values of $\mu A$. On the other hand, invoking the condition

\begin{equation}
\mu A = 2(n+1)
\end{equation}
will terminate the series at a finite value of $n$. So it is of some interest that a single series permits both a classical continuum solution and an infinite number of discrete solutions, as well. This is the direct effect of the gravity field on the Proca interaction.

So when $\mu$ and $A$ are both positive, the quantity $\mu A$ can be quantized. The mass of a particle that exhibits the strong or color force corresponding to the positive exponential solution, therefore, must be proportional to an integer multiple of the inverse of $\mu$, which is associated with the fractional rate of change of the Proca force strength.

As an alternative, a second substitution can be used as a trial solution. However, substituting  $g(z) = h_{1'}(z) e^{-z}$, where $z = \mu y$, yields a recursion relation of

\begin{equation}
a_{n+1} = \left(\frac{\mu A + 2(n+1)}{(n+1)(n+2)}\right)a_n
\end{equation}
In this case there are no eigenvalue conditions that make sense unless $A <0$, which would correspond to a negative mass. Like the previous case, in the limit as $A$ goes to zero, the series gives a standard Proca solution, $u_{1'} = \sinh z/ z$. Is this solution distinct from the previous solution, or the same solution in a different form? A numerical calculation indicates that although $h_1 \ne h_{1'}$, $u_1 = h_1e^z/z = h_{1'}e^{-z}/z = u_{1'}$ identically. Because these functions are smooth, the eigenfunctions must also be the same for both choices, with a trivial overall change in sign that can be absorbed by the integration constant, $a_0$. For very small $\mu A $, the continuum solution looks like $\sinh z/z$. The eigenvalue solutions are all polynomials multiplied by $e^z/z$.

\subsection{ Independent Solution}

The method of Frobenius \cite{Schaums} gives a second independent solution depending on the difference of the two roots of the characteristic polynomial. Here, the roots differ by an integer, so a second solution can be found by positing 

\begin{equation}
h_2 = b_{-1} (\ln z) \sum_0^\infty a_n z^{n + \lambda_1}  +   \sum_0^\infty b_n z^{n+ \lambda_2}
\end{equation}
where again $\lambda_1 = 1$ and $\lambda_2 = 0$. The recursion relationship can again be found by straightforward calculation:

\begin{equation}
b_0 = \frac{b_{-1} a_0}{\mu A}
\end{equation}

\begin{equation}
b_{n+2} = \left(\frac{\mu A - 2(n+1)}{(n+1)(n+2)}\right) b_{n+1} - \left(\frac{(2n+3)\mu A - 2(n+1)^2}{(n+1)^2 (n+2)^2} \right)a_n
\end{equation}
with $b_1$ arbitrary. It's important to notice that terms with a surviving $\ln z$ factor, taken together, sum to zero. The first power series in the $a_n$ will naturally converge as before. The power series in the coefficients $b_n$ will also converge due to the factors involving $n$ in the denominators of the two expressions.   As before, there are numerous quantized solutions when $\mu A = 2(n+1)$. In general, the $b_n$ power series doesn't terminate, it just changes form, with a pair of truncated series involving the $a_n$ terms. Further quantized solutions are possible for the case in which $(2n+3)\mu A = 2(n+1)^2$. There are no solutions involving both this new condition and the original condition simultaneously, because that can be shown to require negative values of $n$.

\section{Discussion}
The minimally-coupled Einstein-Proca equations, to zero order which corresponds to the Proca solved on a Schwarzschild background, show that in spacetime, a simple potential from a second order partial differential equation can result in a surprisingly rich family of solutions.  Physically, these solutions must correspond to regions where gravity is sufficiently weak that the Schwarzschild solution is a good approximation. That is likely true for subnuclear length scales less than a femtometer, but much larger than the Planck length of about $10^{-35} \; \mathrm{m}$. It's a matter of conjecture that the strength of gravitation gradually increases at smaller scales. At the scale of quarkonia and larger, these solutions are surely valid. Although it's speculation, they may be valid for several orders of magnitude smaller than that.

An interesting question is the existence of quantizations of $\mu A$ and what it might imply. Naturally, $A = 2mG/c^2$, and as long as particles can be considered point-like, the quantized solutions imply $\mu m = c^2(n+1)/G$. For a given value of $\mu$
 there are an infinite number of such Proca states, and presumably a given particle could transit from one to another. The set of solutions with negative mass and factors of $e^{-\mu z}$ would give another family of negative mass particles. Negative mass has been invoked in physics, for example in renormalization, and occasionally studied theoretically, but there isn't any experimental evidence and it appears to be a theoretical artifice.
 
 The second independent solution is interesting in that it features a natural logarithm, another function sometimes used in the study of quarkonia. Linear combinations of these two solutions yield a wealth of potentials. The quantization condition for the first potential will terminate the logarithm term of the second potential, but not necessarily the series involving $b_n$, in part due to the fact $b_1$ is arbitrary, but also because the second term is a multiple of $a_n$, preventing the simultaneous vanishing of both terms of the $b_n$ coefficients. Finally, the second solution would permit a series of quantized rational values of $\mu A$.
 
 In principle it may be possible to determine if these functions describe physical phenomena at some energy level. If so, they might also affect the final state of collapsed matter such as neutron or quark stars. It is at least conceivable that experimental or observational evidence could be found that would validate or refute the existence of these potentials in nature.

\section{Conclusions}

It's clear that the Proca equation against a Schwarzschild background admits quite different potentials from the flat space case. The first independent solution gives both a continuum solution and eigenvalue solutions that quantize $\mu A$, terminating a power series in $\mu(r-A)$ that multiplies the exponential expression $e^{\mu (r-A)}/\mu(r-A)$. A second, independent solution involves a natural log, the first power series, and a second power series, again multiplied by $e^{\mu (r-A)}/\mu(r-A)$.  So in a curved spacetime, the actual Proca potentials take forms that vary significantly from those in flat space. It would be of interest exploring some of the consequences of these potentials, which could conceivably pertain to an intermediate-strength gravity regime. Further,similar calculations can be carried out for the oscillatory Proca equations, which correspond to the opposite sign in the mass term. The oscillatory Proca potentials (like the hyperbolic Proca potentials) can give good fits to bound states of quarkonia in simple Bohr models. \cite{VuilleBohr} Another area of interest to pursue would be analytic continuation to transform such solutions into cosmological solutions, with possible application to the early universe.

\section{Acknowledgements} I would like to thank Jim Ipser of the University of Florida for numerous useful discussions on this work. I would like also to remember the late Steve Detweiler of the University of Florida for his many years of friendship and encouragement. Steve, a member of the Academy of Sciences, died while jogging shortly after his retirement. Finally, I am grateful to Doren Poenaru for encouraging my work on the Einstein-Proca equations.

\end{document}